\documentstyle[preprint,aps]{revtex}
\tightenlines

\begin{document}

\title{Amplitude Analysis}
\author{Ron Workman}
\address{Department of Physics, Virginia Tech,
Blacksburg, VA 24061, USA}

\maketitle
\begin{abstract}
Even if a `complete set' of experimental observables were
measured for the elastic scattering or photo/electroproduction
of pseudoscalar mesons, ambiguities would remain in the
extracted partial-wave and isospin decomposed amplitudes.
As these problems are not widely understood, the present work
outlines the way model-dependence enters into analyses of data
from both hadronic and electromagnetic facilities.
\end{abstract}

\section{Introduction}

In order to understand the significance of amplitude analysis,
one must first recognize that there is generally a gap between
the quantities `predicted' through models and the observables
which are actually `measured' in experiments. The experimental
quantities are generally cross sections, often with polarizations
fixed or measured for the beam, target and recoiling particles.
In contrast, the `predicted' quantities (for example, the coupling
constants, sigma terms, and resonance properties) typically require
not only the underlying amplitudes, but also extrapolations of these
into unphysical regions and/or separations into background and
resonant contributions.

It is often assumed that at least the first step (data$\to$amplitudes)
in comparisons between theory and experiment could be carried out in a
`model-independent' way, if sufficient observables were available.
This leads to a working definition of amplitude analysis:

\smallskip

``Amplitude analysis is a model-independent determination of
the (helicity/transversity)
amplitudes using only experimental data and the relations between
amplitudes and observables, resulting in a set of amplitudes at
discrete energies and angles.''

\smallskip

The usual goal in this type of analysis is the smallest set of
observables removing all ambiguities in the extracted amplitudes,
apart from an overall phase. Unfortunately, one rarely has the
required set of observables. As a result, further theoretical
input is generally required. In the following we will first
review some of the most important (formal) results of these
studies. We will then consider how relevant these results are
for the N$^*$ program. Finally, since most analyses depart from
the ideal defined above, we will consider how model-dependence
enters into and effects a range of more standard analyses.

\section{Some Formal Results}
\label{secform}

Most of the formal work on amplitude (and partial-wave) analysis has
been confined to spinless and spin-0$-$spin-1/2 cases. A good source of
references is the Introduction of H\"ohler's `bible' on the
subject\cite{Hoehler}.  Another very readable source is the review of
Bowcock and Burkhardt\cite{Bowcock}. More mathematical discussions are
contained, for example,
in the papers of Sabba Stefanescu\cite{Sabba} and the book of
Chadan and Sabatier\cite{Chadan}.

The basic problem is most easily demonstrated in spinless scattering.
Here we have only the cross section
\begin{equation}
{ {d \sigma} \over {d\Omega} } = | f |^2
\end{equation}
which gives the scattering amplitude,
\begin{equation}
f  = | f | e^{i \phi (E , \theta )}
\end{equation}
up to an overall phase. The difficulty comes in determining $\phi (E,\theta)$
with the addition of isospin, analyticity and unitarity constraints.
It appears\cite{comment1}
that this is possible in spinless and spin-0$-$spin-1/2 scattering.
Clearly the task is easier in the forward direction (optical theorem)
and for elastic scattering.

When the scattered particles have spin, there are more amplitudes
to determine and more observables to choose from. If we once again
ignore the overall phase, the object is to pick a set of observables
which fixes all the {\it relative} phases between the amplitudes. The
allowable sets\cite{comment2}
have been determined by Dean and Lee\cite{Dean} (for
$\pi N$ scattering), Chiang and Tabakin\cite{Chiang} (for
pseudoscalar meson photoproduction), and Dmitrasinovic, Donnelly and
Gross\cite{Gross} (for scalar and pseudoscalar electroproduction).

\section{Are we asking the right question?}

Suppose now that we have been given a complete set of observables for
a particular reaction, and have constructed the helicity amplitudes
up to an overall phase. Is this sufficient to determine all of the
contributing N$^*$ states? For two important reasons, the answer is
negative. What we really want are the partial-wave amplitudes
for each isospin state. Once the (model-dependent) separation of
resonant and background contributions has been carried out, we can
compare with model predictions. Unfortunately, the unknown overall
phase is an important element in both the partial-wave and isospin
decompositions.

To illustrate the first point, we can return to the simple spinless
case, where the full amplitude
\begin{equation}
f \; = \; \sum_l (2l+1) f_l (E) P_l (x)
\end{equation}
may be written as an infinite sum of partial-wave amplitudes. Here
$x$ denotes $\cos (\theta )$. Clearly, if $f$ has a phase which is
an unknown function of $\theta$, this relation cannot be inverted
to give unique values of $f_l (x)$.

The effect of this overall phase on isospin decompositions is also
serious, but has been discussed less often. This may be due to the
fact that, in $\pi N$ scattering, the potential ambiguities can be
easily removed. Suppose, in the $\pi N$ case, that we
have constructed amplitudes for elastic $\pi^{\pm} p$ scattering and
charge-exchange ($\pi^- p\to \pi^0 n$) scattering. In principle, the
amplitudes for each reaction have a {\it different} overall phase.
Therefore we cannot simply combine these quantities, using the usual
relations, in order to obtain isospin amplitudes. However, we know
one more piece of information. The amplitudes for the charge states
satisfy a triangle relation
\begin{equation}
A^- \; + \; \sqrt{2} A^0 \; = \; A^+
\end{equation}
which, since we know the lengths of all 3 sides, fixes the relative
angles.

Unfortunately, this method doesn't work in all cases. As an example,
consider pion photoproduction. Here there are 4 different amplitudes
for the production of $\pi^0 p$, $\pi^+ n$, $\pi^- p$, and $\pi^0 n$.
These can be constructed from 3 isospin amplitudes. In this case, we
also have a relation
\begin{equation}
\sqrt{2} A^{\pi^0 n} + A^{\pi^- p} + A^{\pi^+ n} = \sqrt{2} A^{\pi^0 p}
\end{equation}
between the charge states. However, knowing the lengths of the 4
sides is now {\it not} sufficient to determine the relative angles.
This problem remains even if the overall phases
have been fixed for 2 of the reactions.

Some readers may at this point be wondering how an `unmeasureable'
phase could be so important. It should first be noted that this phase
is not equivalent to the unmeasureable phase of a wavefunction in
quantum mechanics. Instead, this is the relative phase between the
scattered and unscattered waves in the familiar relation
\begin{equation}
\psi \sim e^{i\vec k \cdot \vec x} + f(\theta )
{ {e^{ikr} }\over {r} } \;\;\; {\rm as} \;\;\; r\to \infty .
\end{equation}
It is actually possible to see effects of such phases via
multiple scattering\cite{Hoehler,Bowcock} and to (in principle)
measure some of them using the
Hanbury-Brown$-$Twiss method\cite{comment3}. In practice,
however, some theoretical input is necessary to fix this phase.

\section{What is generally done}

Given the problems with `pure' amplitude analysis, it
is not surprising that alternate approaches have been employed in the
study of N$^*$ physics. These include restricted multipole analyses,
the use of dispersion relations, direct model fits to data, and fits
based on Breit-Wigner plus background contributions.

Before discussing
these, however, we should mention an approach to the amplitude
analysis problem which has been applied to elastic $\pi d$ scattering.
In this work\cite{Garcilazo} there were insufficient observables for
a model-independent amplitude analysis. As a result, amplitudes were given
with 3 different levels of model-dependence. First, it was shown that
certain combinations of amplitudes could be extracted directly from the existing
data. Models then provided the additional information necessary to complete
a full amplitude analysis. Finally, a model was used to fix the overall
phase, thus allowing a partial-wave analysis. An analysis of this type in
the N$^*$ arena would also be interesting, particularly as more polarization
data become available.

One way of avoiding the overall phase problem is to fit multipoles
directly. In the simplest variant of this method,
the partial-wave series is cut off after a few terms.
Examples include analyses near threshold and
over the first resonance region. Clearly this method won't work if the
neglected terms sum up to a sizable contribution\cite{Donnachie}.
Often the higher
waves are assumed to be well approximated by the Born terms. A particularly
interesting study of this kind is the Grushin\cite{Grushin} fit. This
analysis was carried out over the first resonance without the help of
pion nucleon phases (which give the multipole phases via Watson's theorem).
Here the overall $\pi^+ n$ phase was determined through interference with
the (real) Born terms. A relation between the $\pi^0 p$ and $\pi^+ n$
multipoles was then used to fix the overall $\pi^0 p$ phase.
Such an analysis, free from $\pi N$ input, can form the basis for a
relatively unbiased determination of the $\Delta^+$ resonance
position\cite{delta}.

Dispersion relations, applied either to the invariant or partial-wave
amplitudes give the least model-dependent way to supplement experimental
data. Analyticity, particularly for fixed-$t$, has been an important element
in studies of uniqueness for spinless and spin-0$-$spin-1/2
scattering. These constraints are particularly useful in elastic scattering
as they give the real parts of forward amplitudes, once the imaginary
parts have been determined via the optical theorem and total cross sections.
To the author's knowledge, no formal studies (analogous to those carried out
by Sabba Stefanescu for $\pi N$ scattering), have determined the minimal
theoretical input required to fix all the phases for pseudoscalar meson
photo and electroproduction. Here, even the appropriate question is less
obvious, as there are many more observables which could be
measured.

If the analytic properties of the amplitudes are assumed in advance
(for example,
in Breit-Wigner plus background or model-based fits), we are able to obtain
results from minimal sets of data. A complete set of experiments is not
required, and the results are obviously model dependent.
Here the ambiguity lies in the choice of
model, which must necessarily give only an approximation to the true analytic
structure. This approach can also yield useful information, particularly when
it fails. Failure within a model (hopefully one which builds in the most
important constraints from analyticity and unitarity) indicates a missing
element.

\section{Implications for CLAS data analysis}

At present, our knowledge of the N$^*$ spectrum comes from analyses of data
from decades of measurements at numerous laboratories, carried out
using a variety of techniques (each having different inherent systematic
errors). The promise of CLAS data has been a set of measurements with high
precision and linked systematic errors. This would greatly reduce one of the
most difficult problems in amplitude and partial-wave analyses. It is
therefore interesting to consider how much information we can obtain from
CLAS data alone.

If one is able to control the beam and target polarization, while measuring
in both $\theta$ and $\phi$, it has been shown\cite{BDS} that one can obtain
the type-$S$ (cross section and single-polarization) and $BT$ (beam-target
double-polarization) data, including the recoil polarization $P$. In
pion photo- or electroproduction, for example, this leaves only one relative
phase undetermined, apart from the overall phase. How one could obtain $P$
without recoil-polarization measurements is also evident in the relation
\begin{equation}
FG - EH = P - T\Sigma
\end{equation}
between the type-$S$ and $BT$ observables\cite{BDS}.

The problem, from an `amplitude analysis' point of view, is that this remaining
relative phase requires further double-polarization measurement involving
recoil polarization. As CLAS was not designed for such measurements, it might
seem that non-CLAS data would be required to complete a `model-independent'
analysis. However, as we have argued above, in order to extract partial-wave
amplitudes, the overall phase must be fixed. By fixing this overall phase,
relative phases are also restricted. It seems reasonable to `conjecture' that
any physical input sufficient to fix this overall phase will simultaneously
remove (at least) the one relative phase undetermined from type-$S$ and $BT$
experiments. The remaining question then has a more theoretical nature.
What (minimal) theory input is required to fix the overall phase in photo-
electroproduction reactions?

The author thanks B.M.K. Nefkens for sparking his interest in these and
related questions.
This work was supported by U.S. Department of Energy grant No.
DE-FG02-97ER41038.


\begin{thebibliography}{9}
\bibitem{comment2}
Note that forward and backward scattering are special cases for the
amplitude reconstruction problem. For example, in $\pi N$ scattering
and photoproduction, the spin observables are either zero or $\pm 1$
at these angles. Therefore we need only a cross section measurement
to determine these (essentially degenerate) transversity amplitudes
up to an overall phase.
\end{thebibliography}
\end{document}